\documentclass[12pt]{article}

\usepackage[margin=1in]{geometry}
\usepackage[T1]{fontenc}
\usepackage[utf8]{inputenc}
\usepackage{lmodern}
\usepackage{setspace}
\usepackage{booktabs}
\usepackage{tabularx}
\usepackage{longtable}
\usepackage{array}
\usepackage{amsmath,amssymb}
\usepackage{graphicx}
\usepackage{subcaption}

\usepackage{csquotes}
\usepackage[
    backend=biber,
    style=apa,
    natbib=true
]{biblatex}

\usepackage[hidelinks]{hyperref}

\usepackage{xcolor}
\usepackage{enumitem}

\graphicspath{{figures/}}

\title{\textbf{Meeting the Coming Wave: The Emerging Politics of AI and Work across 33 Parliaments}}
\author{
Juliana Chueri\thanks{Department of Political Science and Public Administration, Vrije Universiteit Amsterdam. Email: \texttt{j.chueri@vu.nl}.}
\and
Petter Törnberg\thanks{Institute for Logic, Language and Computation, University of Amsterdam. Email: \texttt{petter.tornberg@uva.nl}.}
}
\date{}

\begin{document}
\maketitle

\begin{center}
\end{center}

\begin{abstract}
\noindent A new politics of artificial intelligence and work is taking shape across party systems, but comparative politics has yet to map it. Using 1,514,950 parliamentary speeches from 33 parliaments (2023-2026), we show this politics follows a different logic than political economy expects. Research anticipates that technological disruption generates demands for compensation; instead, compensation accounts for just 2.3\% of response-frame mentions, while enablement and investment dominate (55.2\%), regulation and restriction follow (21.8\%), and training (20.6\%) appears at similar rates across families. Parties disagree instead over what AI means for work and how far this technology should be restrained. The mainstream and radical right support unrestricted enablement; the left is critical but divided on remedy. Social democrats stay adoption-oriented; greens split evenly. The radical left is the clearest force for restriction. The AI conflict thus concerns not compensation after disruption, but whether politics should enable technological change or govern its trajectory.
\end{abstract}

\textbf{Keywords:} artificial intelligence; automation; labor markets; party competition; parliamentary debates; future of work

\newpage
\section{Introduction}
In October 2025, U.S. Senator Bernie Sanders released a report warning that artificial intelligence and automation could eliminate nearly 100 million American jobs over the next decade \citep{sanders2025bigtecholigarchs}. A few months later, Sanders and Alexandria Ocasio-Cortez introduced legislation to impose a federal moratorium on new AI data centers \citep{sanders2026aiDataCenterMoratoriumAct}. In Europe, green and left parliamentarians pressed for stronger protections against algorithmic management at work, including rights to information, explanation, and human oversight over automated workplace decisions \citep{bundestag2025digitaleSouveraenitaet}. Other governments and parties have responded in a very different register. Singapore has made AI adoption a centerpiece of its economic and workforce strategy \citep{edb2024singaporesNationalAIStrategy}, and parties around the world echo its language of investment, competitiveness, and modernization. These positions suggest that parties are beginning to construct very different political meanings around AI, work, and technological change. A new politics of AI and work is taking shape---quickly and unevenly across party systems---and comparative politics has yet to map it.

Artificial intelligence is no longer only a speculative future, projected by economists or imagined in science fiction; it is emerging as a new arena of party conflict. Generative systems are entering occupations that long seemed sheltered from automation \citep{felten2021occupational, FeltenRajSeamans2023}, raising concerns not only about job loss but also about deskilling, workplace surveillance, algorithmic management, bargaining power, and the allocation of productivity gains \citep{GoosSavona2024, BaioccoFernandezMaciasRaniPesole2022, Casilli2025, LauterbachToberKunzeBusemeyer2023}. While the actual consequences remain uncertain, the politics of AI is already taking shape---and the question for comparative politics is therefore immediate: how are parties turning AI's consequences for work into political conflict?

A prominent strand of political economy offers a clear expectation. Automation scholarship on parties and policy typically begins from labor-market risk: technological change creates winners and losers, and politics responds by debating compensation, social insurance, redistribution, or retraining \citep{ThewissenRueda2019, KurerHausermann2022, BusemeyerGandenbergerKnotzTober2023, BusemeyerTober2023}. This literature has generated important insights into the political consequences of technological disruption. Yet it captures only one side of the politics of technology: its aftermath. By beginning with exposure and its resulting risks, it tends to enter the political process only after the direction of technological change and its consequences for workers have already been established.

The emerging public debate around AI, however, hints at an earlier stage of technological politics: a politics of technology before its consequences have been settled. Demand-side research suggests that citizens hold preferences not only over compensation and adaptation, but over the pace, direction, restriction, and regulation of technological adoption itself \citep{heinrich2025self, BicchiKuoGallego2025, KuoBurgisserGallegoHausermann2026, HaslbergerGingrichBhatia2025, Chueri2026}. This raises the possibility of a politics that the literature has largely overlooked: one that treats the impact of technology itself, not just its consequences, as a matter of political choice. Whether AI replaces workers or augments them, intensifies managerial control or expands worker capacity, and concentrates or distributes its gains depends on the institutions and decisions governing how technologies are developed, adopted, and used. Politics, in other words, could do more than help workers adjust to technological change or compensate those harmed by it: it could shape the conditions under which disruption occurs and, with them, who eventually wins and loses.

Whether this earlier-stage politics actually materializes on the supply side, however, remains an open question. We still know little about how AI and work is politically constructed as it enters parliamentary debate: whether parties treat it as a problem of compensation for displaced workers, a problem of skills and adaptation, an opportunity for investment and growth, or a problem of control over deployment.

This paper examines the emerging politics of AI and work through a new corpus of 1,514,950 substantive parliamentary speeches from 33 national parliaments between 2023 and 2026. The cases span wealthy and middle-income economies, parliamentary and presidential systems, different welfare-state traditions, party systems, and state strategies toward digital transformation. We combine multilingual retrieval with validated large-language-model coding to identify AI-work speeches, diagnostic frames, and response frames \citep{SnowBenford1988,BenfordSnow2000}. This allows us to distinguish how parties diagnose AI's consequences for work---as threats or opportunities---from the forms of political action they attach to these diagnoses: regulation and restriction, enablement and investment, training, or compensation. We then estimate hierarchical models with party and country random intercepts and year fixed effects to examine how this emerging issue is structured across party families.

The findings show that this politics has emerged fast, but far from evenly. AI attention and AI-work salience both rise between 2023 and 2026, yet even by 2026 AI-work speech remains below one percent of all parliamentary speeches. Political debate has a strongly adoption-oriented center of gravity, but also a clear and asymmetric line of contestation: enablement and investment account for 55.2 percent of broad response-frame mentions, regulation and restriction for 21.8 percent, training for 20.6 percent, and compensation for just 2.3 percent. The politics anticipated by the demand-side literature, in other words, is not merely a latent possibility: it is already visible in party supply, well before AI's labor-market consequences have been settled. We refer to this as the \textit{politics of adoption}, distinct from the \textit{politics of compensation} that has been the primary focus of the existing literature.

Conservative, liberal, and Christian-democratic parties overwhelmingly frame AI as an opportunity and promote adoption and investment. The radical right aligns with them: although automation exposure has repeatedly been linked to radical-right support, this party family is not politicizing AI as a threat to workers. Social democrats sit between this adoption-oriented right and a more contestatory left: they recognize AI as both threat and opportunity in roughly equal measure, yet their aggregate response still leans toward enablement rather than restricting deployment. The radical left constitutes the clearest force of contestation: its dominant response is regulation and restriction, reflecting a broader diagnosis of AI as a threat to workers. Greens lean the same way on diagnosis, though less decisively, and split evenly between regulation and enablement in response. Training cuts across this divide, invoked at comparable rates across every party family---serving adoption for some, worker adaptation for others. Compensation shows the opposite pattern: it remains rare throughout the party system, and what little compensation politics exists is concentrated almost entirely on the left. A further gradient runs through institutional position. Net of party family, country, and year, government parties are more likely to frame AI as an opportunity and to favor enablement, while opposition parties lean more toward threat and regulation.

Beyond confirming that this politics of adoption exists, the analysis speaks to a further debate in comparative political economy: which parties drive it. The radical left, and to a lesser degree the greens, are the principal parliamentary actors making AI's labor-market trajectory politically contestable, while most other party families treat adoption as a process to be enabled and managed rather than governed. This includes the radical right: research shows that workers exposed to automation risk are more likely to support radical-right parties \citep{Knotz2025, ImMayerPalierRovny2019, AnelliColantoneStanig2021, GallegoKurer2022}, yet this electoral link does not translate into a distinct party-level position in relation to AI and work. Radical-right parties do not develop a separate stance on AI and automation, aligning instead with the adoption-oriented right. The emerging conflict over AI is thus not simply about who should be compensated for technological change: it is about whether the direction and consequences of technological change should themselves be subject to collective choice.

\section{Shaping technological change: AI and the politics of work}
\label{sec:beyondcomp}
What kind of politics should we expect AI's impact on work to generate? The existing literature on technology and labor-market politics offers a clear starting point, built around occupational risk. In this view, politics primarily responds to technological impact: automation displaces workers, depresses wages, or hollows out occupations, and only then does political conflict begin, organized around how to manage the resulting losses through compensation, social protection, redistribution, and retraining \citep{ThewissenRueda2019}. Later work has refined this account, showing that citizens distinguish among policy instruments---when facing automation risk, workers tend to support compensation policies, but not necessarily social investment \citep{KurerHausermann2022, BusemeyerGandenbergerKnotzTober2023, BusemeyerTober2023}. What remains constant across this literature, however, is its starting point: politics enters the story downstream of technology, once exposure has occurred and its consequences for particular workers are becoming visible.

Two developments suggest that this may not be the whole picture. First, demand-side research on AI shows that technological risk does not generate only demands for compensation: citizens also express preferences over slowing technological change, regulating AI, protecting work, accelerating digitalization, and investing in adaptation and skills, depending on their perceptions of individual and societal gains \citep{GallegoKuoManzanoFernandezAlbertos2022, BicchiKuoGallego2025, heinrich2025self, Chueri2026, KuoBurgisserGallegoHausermann2026, HaslbergerGingrichBhatia2025}. Citizens, in other words, hold preferences not only over how losses should be handled after the fact, but over the pace, direction, and terms of technological change itself. Second, AI's effects on work extend beyond displacement to the organization of supervision, monitoring, evaluation, expertise, and authority at work \citep{AcemogluRestrepo2020, AcemogluAutorJohnson2023, GoosSavona2024, BaioccoFernandezMaciasRaniPesole2022}. A worker whose tasks are monitored by an algorithm, whose professional judgment is displaced by a model, or whose output is priced against automated alternatives experiences a shift in economic power well before any unemployment spell or benefit claim would register in a compensation-focused account. Together, these developments suggest a political space that extends beyond the treatment of losses after disruption---one that existing supply-side accounts of automation politics, with their focus on compensation, are not built to capture. Whether parties give these preferences political expression, and how they organize the resulting conflict, remains an open question.

Such a space is plausible once we recognize that technologies do not enter labor markets as autonomous shocks with fixed consequences. A broad scholarship on technology has shown how its impact is shaped by institutions, regulation, managerial strategies, labor power, and the broader networks of authority in which they are embedded \citep{Winner1980, MacKenzieWajcman1999, Thelen2004, Rahman2018}. The same technology can displace workers, augment them, intensify surveillance, support collective capacity, or redistribute productivity gains, depending on how it is adopted and governed. Politics can therefore enter before stable groups of winners and losers have formed: public authorities can steer investment, subsidize adoption, and build infrastructure and public-sector capacity to accelerate technological change, or they can regulate workplace use, create rights over automated decisions, strengthen worker voice, and restrict particular applications to condition it.

This gives the citizen preferences and workplace dynamics documented above a natural political counterpart. If technological consequences are not fixed but shaped by how change is governed, then the central question facing public authority is not only how to compensate for disruption, but whether to accelerate it or constrain it. 

Four broad responses capture this wider space \citep{fan2026slowing,hunter2026political}. Parties may seek to \emph{enable adoption} through investment, infrastructure, firm support, public-sector capacity, and technological competitiveness. They may emphasize \emph{training}, preparing workers and organizations for anticipated change. They may seek to \emph{regulate or restrict} technological change by conditioning deployment, strengthening worker voice, creating rights and safeguards, limiting harmful applications, or slowing labor-replacing adoption. Or they may offer \emph{compensation} through income support, social protection, redistribution, reduced working time, or basic income.

These responses enter at different points in the process. Compensation absorbs losses; training adapts workers and organizations; enablement expands capacity and accelerates adoption; regulation and restriction shape the pace and terms of deployment. They can coexist, and they do not necessarily form a simple more-versus-less-state continuum. Governments may simultaneously subsidize AI, train workers, and impose workplace safeguards. The distinction concerns the purpose for which political authority is mobilized: repairing consequences, preparing society to accommodate change, accelerating it, or governing its trajectory.

\begin{table}[t]
\centering
\caption{Response frames and the political questions they answer}
\label{tab:frames}
\begin{tabularx}{\textwidth}{@{}p{0.21\textwidth}p{0.28\textwidth}p{0.23\textwidth}X@{}}
\toprule
Response frame & Political question & Main mechanism & Examples \\
\midrule
Compensation & How should losses or reduced labor demand be absorbed? & Income protection and redistribution & unemployment insurance, social protection, reduced working time, basic income \\
Training & How should workers and organizations adapt? & Human-capital adjustment & reskilling, lifelong learning, AI literacy \\
Enablement and investment & How should adoption and capacity be expanded? & Investment and capacity building & firm support, infrastructure, public AI strategy \\
Regulation and restriction & How should the pace and terms of deployment be governed? & Conditioning technological change & worker voice, safeguards, limits on surveillance, moratoria, robot taxes \\
\bottomrule
\end{tabularx}
\end{table}

An issue space, however, identifies positions that parties \emph{could} occupy rather than positions they necessarily will. Some responses may dominate while others remain largely vacant; some may spread across the party system while others become concentrated in one political family. These patterns are themselves evidence about how a new technological conflict is being constructed.

We capture that construction by distinguishing between \textit{diagnostic} and \textit{prognostic} framing \citep{SnowBenford1988,BenfordSnow2000}. Diagnostic frames define what AI means for work. \emph{Threat} includes claims concerning displacement, job insecurity, deteriorating job quality, weaker autonomy, inequality, and difficulties navigating the skills transition; \emph{opportunity} includes claims about productivity, worker augmentation, better jobs, public or private capacity, competitiveness, and modernization \citep{hunter2026political}. Response frames capture the political action attached to these diagnoses. \emph{Regulation and restriction} includes rights, safeguards, worker voice, collective bargaining over algorithmic systems, moratoria, bans, and taxes intended to discipline labor-replacing adoption \citep{Ehret2022,Chueri2026,BicchiKuoGallego2025,AcemogluLensman2024}. \emph{Enablement and investment} covers adoption support, AI capacity, infrastructure, and innovation \citep{hunter2026political,KuoBurgisserGallegoHausermann2026}; \emph{training} covers reskilling, education, and lifelong learning \citep{KurerHausermann2022,HaslbergerGingrichBhatia2025,fan2026slowing}; and \emph{compensation} covers income protection, redistribution, reduced working time, and basic income \citep{dermont2020automation,BusemeyerGandenbergerKnotzTober2023,BusemeyerTober2023,fan2024does}.

Separating diagnosis from response allows us to locate the emerging conflict more precisely. Parties may recognize similar risks and propose different remedies, or their differences may begin earlier, in what they make politically visible as a threat or an opportunity. The supply-side question is therefore both who talks about AI and work and how parties construct the technology once they do.

\subsection*{Party supply and the construction of technological change}

New technologies enter party competition through political vocabularies that already exist. AI can be linked to job loss, workplace control, labor shortages, national competitiveness, skills, corporate power, or public-sector modernization, and parties foreground such connections when they fit existing constituencies, reputations, and programmatic commitments \citep{BudgeFarlie1983, Petrocik1996, Meguid2005}. Research on digitalization likewise shows that parties often avoid technological issues when ownership is unclear, but interpret them through ideology, competence, and strategic issue expansion once they engage \citep{KonigWenzelburger2019, BressanelliBuzzelliMuccilliSacchi2025, Guglielmo2025}. AI-work politics should therefore be structured less by the technology in the abstract than by the political vocabularies parties already possess.

AI enters party systems already transformed by postindustrialization. The decline of the industrial working-class vote, the expansion of higher education, the growth of service employment, and the reconfiguration of welfare coalitions have made labor-market politics less reducible to a simple class cleavage \citep{kitschelt1990left, gidron2022many}. Parties now compete in a pluralized issue space in which economic protection, social investment, cultural liberalism, national identity, and producer interests are bundled differently across party families. For social democrats, it creates a tension between defending workers with varying levels of employment protection and promising competent modernization \citep{Rueda2005, GingrichHausermann2015}. For radical-right parties, it can be folded into economic insecurity and anti-elite rhetoric, but it also competes with narratives of digital nationalism and technological sovereignty \citep{goode2021artificial}.

These differences are likely to shape both the diagnoses parties offer and the responses they attach to AI-related labor-market change. Left, green, and social-democratic parties have reputational advantages on worker protection, equality, public regulation, and the limits of employer authority \citep {goldmann2025new}. They are therefore likely to diagnose AI as a threat to employment, job quality, skills, and equality \citep{hunter2026political}. Yet their response should not be compensation alone. Social-democratic parties increasingly represent coalitions that include secure insiders and middle-class constituencies, not only workers most exposed to displacement \citep{GingrichHausermann2015, Ares2022, AbouChadiHausermannMittereggerMosimannWagner2025}. This should orient them toward managed adaptation: regulation, restriction, worker voice, and training rather than income replacement alone. Radical-left parties, more closely associated with insecure workers and conflictual labor-market claims \citep{Marx2014, Kweon2018}, should be especially likely to combine threat diagnoses with stronger constraints on deployment and support for compensation. Greens' coalition is defined less by labor-market insecurity than by demands for democratic and environmental oversight of technology; this should orient them toward deployment regulation, while their ecological-modernization current leaves them more open than the radical left to selective enablement when adoption serves social or environmental goals \citep{Marquardt2024,Guglielmo2025}.

Liberal, conservative, and Christian-democratic parties approach AI from a different set of reputational resources, as their stronger ties to business, growth, modernization, and competitiveness make opportunity frames more available. They are therefore likely to present AI as a source of productivity, state capacity, innovation, and future employment \citep{goldmann2025new, hunter2026political}. This does not imply an absence of state action, as enablement may consist of active support for acceleration, adoption, and public investment in skills \citep{KuoBurgisserGallegoHausermann2026}. We therefore expect mainstream right and liberal parties to emphasize enablement, investment, and training rather than compensation or restrictive
regulation.


Radical-right parties are the most ambiguous case. Their electorates include workers exposed to labor-market decline and status threat, and automation vulnerability has been linked to radical-right support in several settings \citep{ImMayerPalierRovny2019,AnelliColantoneStanig2021, GallegoKurer2022, Knotz2025}. This creates incentives to mobilize AI-related threat. At the same time, appeals to national interest may push them toward opportunity frames centered on technological sovereignty, competitiveness, and accelerated AI adoption \citep{goode2021artificial}. More generally, radical-right parties have incentives to translate AI-related insecurity through their existing repertoire of national decline, elite betrayal, immigration, and cultural conflict \citep{BorweinBonikowskiLoewen2024}, rather than through labor protection alone. We therefore expect radical-right parties to use threat frames selectively, and to attach those threats less often to worker voice, broad social investment, or institutionalized workplace regulation than the left does \citep{enggist2022radical,chueri2022}.

Institutional position may add a further gradient. Governments are responsible for economic performance, procurement, public-sector modernization, and implementation, giving governing parties stronger incentives to present AI as useful, governable, and compatible with national strategy \citep{KonigWenzelburger2019}. Government parties should therefore be more opportunity- and enablement-oriented than opposition parties, although strong party-family differences would indicate that ideological positioning remains the principal organizing force.

These arguments yield five expectations.

\begin{enumerate}[label=\textbf{E\arabic*.}, leftmargin=2.2em]
\item \textbf{Uneven emergence.} AI-work salience should rise over time but remain modest, with substantial variation across countries and party families while the issue is still forming.
\item \textbf{Regulation over compensation.} Because AI's labor-market consequences remain unsettled, AI-work politics should be organized primarily around whether adoption should be accelerated and enabled or conditioned and restricted, rather than around post-disruption welfare responses. Regulation and restriction, and enablement and investment, should therefore be more prominent than compensation.
\item \textbf{Diagnostic sorting.} Left, green, and social-democratic parties should diagnose AI more often as a labor threat; liberal, conservative, and Christian-democratic parties should diagnose it more often as an opportunity. If party positioning follows available ideological repertoire rather than electorate exposure, radical-right parties should pattern with the mainstream right rather than showing a distinct threat orientation.
\item \textbf{Response sorting.} Left, green, and social-democratic parties should attach AI-related labor threats primarily to regulation, restriction, worker voice, and training; because postindustrialization has diversified left constituencies beyond a single working-class base, this response should not be uniform across the left---parties anchored in broader, insider-inclusive coalitions (social democrats) should lean more toward enablement than parties anchored more narrowly in oversight-oriented or conflictual labor claims (greens, the radical left). Liberal, conservative, and Christian-democratic parties should emphasize enablement, investment, and training.
\item \textbf{Government moderation.} Government parties face stronger incentives than opposition parties to present AI as governable and compatible with national strategy \citep{KonigWenzelburger2019}. Government speakers should therefore be more opportunity- and enablement-oriented than opposition speakers, net of party family, country, and year, though large party-family differences would indicate that ideology remains the stronger organizing force.
\end{enumerate}

\section{Data and methods}

\subsection*{Corpus}

Parliamentary speech provides a valuable record of an issue while its political boundaries are still being formed. AI's labor-market consequences remain unsettled, yet political actors are already publicly defining what the technology means for workers, whom it threatens or benefits, and what forms of action it requires.

We assembled a new corpus from official parliamentary sources covering 33 national parliaments between January 2023 and April 2026: Argentina, Australia, Austria, Belgium, Brazil, Canada, Chile, Croatia, Czechia, Denmark, Estonia, Finland, France, Germany, India, Ireland, Italy, Japan, Kenya, Latvia, the Netherlands, New Zealand, Norway, Poland, Singapore, Slovenia, South Africa, Spain, Sweden, Switzerland, Taiwan, the United Kingdom, and the United States. The analytic corpus contains 1,514,950 substantive speech turns by parliamentarians and government speakers, excluding procedural interventions and chair turns.

Cases were included where official parliamentary records could be recovered at the level of individual interventions and linked to basic speech metadata. The sample is therefore not a probability sample of legislatures, but a geographically and institutionally diverse set of countries with recoverable speech-level records during the first generative-AI wave. It spans wealthy and middle-income economies, parliamentary and presidential systems, and diverse party systems. The observation period begins in 2023, after the public diffusion of generative AI created a new political environment, and ends when data collection was completed in April 2026; all 2026 estimates therefore cover January through April.

The corpus was assembled with country-specific parsers drawing on official APIs, XML feeds, HTML reports, Hansard-style transcripts, and PDF publications. To maximize comparability, we focus on lower-house plenary debate, or the national unicameral chamber where applicable. All sources were normalized into a common schema containing country, date, sitting, debate, speaker, party, text, and source provenance, with one row representing one substantive intervention. See Appendix for further details.

Party affiliation was assigned at the date of each speech using transcript metadata where available and otherwise official rosters, member APIs, and historical directories. This produced 936 country-party units, which remain distinct across countries throughout the analysis. For comparative analysis, parties were harmonized into seven families: radical right, conservative, liberal, Christian democratic, social democratic, green, and radical left. Corpus labels were linked to Party Facts where possible and supplemented with country-specific documentation. We coded conservatively, excluding parties that could not be mapped with sufficient confidence rather than forcing them into the comparative scheme. Party families are identified for 94.7 percent of AI-work speeches, and 85.5 percent fall within the seven families analyzed below. 

\subsection*{Identifying AI-work debate}

AI-work debate is rare, multilingual, and substantively heterogeneous, so we use a two-stage measurement strategy that prioritizes recall at retrieval and precision at classification.

First, multilingual dictionaries covering artificial intelligence, automation, and related algorithmic technologies retrieve a deliberately broad candidate set. In a rare-issue setting, false positives can be removed downstream, whereas speeches missed at retrieval cannot be recovered. This stage yields 19,411 candidate speeches, including many AI speeches unrelated to work.

Second, candidate speeches are classified for AI-work relevance. A speech is \emph{directly relevant} when AI, automation, or algorithmic systems are explicitly connected to employment, tasks, wages, skills, job quality, workplace control, labor-market institutions, or worker protection. It is \emph{indirectly relevant} when the connection is broader, for example through occupational transition, productivity, or public-sector capacity. Adjacent and irrelevant speeches are excluded. This produces 5,317 AI-work speeches, of which 3,076 are directly relevant. All 5,317 form the broad analytic sample; we repeat the central response analyses on the more restrictive direct-only sample. 

\subsection*{Theory-driven frame measurement}

The coding scheme translates the theoretical framework in Section~\ref{sec:beyondcomp} into two sets of measures \citep{SnowBenford1988,BenfordSnow2000}. Diagnostic frames distinguish \emph{threat} from \emph{opportunity}; response frames distinguish \emph{regulation and restriction}, \emph{enablement and investment}, \emph{training}, and \emph{compensation}.

The broad categories were derived from theory rather than discovered from the observed party patterns. Human coders then refined the codebook iteratively across countries, languages, and difficult boundary cases, clarifying recurrent distinctions such as general AI policy versus AI-work policy, worker adaptation versus national skills capacity, regulation versus adoption support, and compensation versus transition assistance. The purpose was to sharpen the empirical boundaries of the theoretical categories.

Every classification is grounded in an explicit claim in the speech. The annotator receives the original-language speech text and the coding instructions, but not the speaker's party family, government status, or the hypotheses tested in the paper. We use zero-shot large-language-model annotation \citep{Tornberg2025,gilardi2023chatgpt}, following established practices for prompt-based text classification \citep{Tornberg_2024}. Coding was conducted with \texttt{gemini-3.1-pro-preview}, accessed through the Google Gemini API and Requesty (\texttt{google/gemini-3.1-pro-preview}) on July 6--7, 2026, with \texttt{temperature = 0} and JSON structured output. See Appendix for full implementation details.

Each relevant speech can receive a primary and secondary diagnostic frame and a primary and secondary response frame. The coding is therefore multi-label: a speech may contain both threat and opportunity, or combine regulation with training. In the main models, a speech counts as containing a frame when that frame appears in either field, and each frame is modeled separately. Predicted threat and opportunity shares, and the four response shares, therefore need not sum to 100 percent. Figures based explicitly on primary frames are mutually exclusive and do sum to 100 percent.

Separating diagnosis from response allows us to examine whether parties differ because they construct AI differently, because they attach different responses to similar diagnoses, or both. The annotation also retains an evidence excerpt and a concise English summary of the AI-work content for auditing and qualitative interpretation; neither enters the statistical models.

\subsection*{Human validation}

We evaluate the measurement pipeline against 728 human-coded observations using two complementary designs. The primary benchmark consists of 538 speeches coded by the lead human coder. Against these labels, the model identifies AI-work relevance with 97.8 percent accuracy, Cohen's $\kappa=0.947$, 96.9 percent precision, and 95.7 percent recall.

Frame performance is evaluated among speeches that both human and model classify as AI-work relevant. Diagnostic coding achieves a micro-F1 of 0.893; response coding achieves a micro-F1 of 0.878. Exact agreement is demanding in a multi-label task because the complete set of assigned frames must match.

A second human coder assessed a deliberately difficult 190-speech stress test enriched for boundary cases, earlier disagreements, and rare response categories. It is not intended to estimate corpus-wide accuracy, but to probe the parts of the codebook most likely to fail. Relevance agreement is consequently lower but still substantial, with 85.8 percent accuracy, Cohen's $\kappa=0.679$, and micro-F1 of 0.894. Among the 114 speeches jointly classified as AI-work relevant, direct/indirect classification reached 96.5 percent accuracy ($\kappa=0.916$), while broad frame distinctions remained robust: diagnosis coding achieved micro-F1 of 0.891 and response coding micro-F1 of 0.828. See Appendix for further details.


\subsection*{Exploratory content analysis}

To examine the substantive content within the validated broad frames, we conduct a supplementary exploratory analysis using the English AI-work summaries retained during annotation. Each speech contributes one mention to every broad frame it receives. Summaries are embedded with \texttt{sentence-transformers/all-mpnet-base-v2} and clustered separately within each frame using $k$-means.

The resulting clusters are used only for interpretation. Labels are based on human review of TF--IDF terms, nearest-centroid cases, and random examples, with substantively overlapping clusters merged. We also reran the solutions across random seeds and adjacent specifications; same-$k$ solutions were highly stable, with mean adjusted Rand indices between 0.889 and 0.928 and mean best-overlap scores between 0.946 and 0.960. Appendix~A reports the full procedure. The clusters do not enter the statistical models or hypothesis tests.

\subsection*{Analytical strategy}

The analysis uses separate specifications for the main party-family comparisons and the government--opposition analysis.

\paragraph{Party-family models.}
The main analyses use grouped-binomial multilevel models estimated on country--party--year cells. For each cell $i$, we observe $y_i$ speeches containing the outcome among $n_i$ eligible speeches and estimate

\begin{equation}
y_i \sim \operatorname{Binomial}(n_i,p_i),
\qquad
\operatorname{logit}(p_i)
=
\alpha
+
\delta_{t[i]}
+
\gamma_{f[i]}
+
u_{c[i]}
+
v_{p[i]},
\end{equation}

where $\delta_t$ are year fixed effects, $\gamma_f$ party-family fixed effects, $u_c$ a country random intercept, and $v_p$ a country-party random intercept. The specification therefore estimates common party-family patterns while accounting for national baselines and repeated observations from the same country-party.

Outcomes differ in their eligible denominator. AI attention is measured as AI candidate speeches among all substantive speeches; AI-work salience as AI-work speeches among all substantive speeches; and conditional work framing as AI-work speeches among AI candidate speeches. Diagnostic models are estimated among speeches with at least one clear diagnosis and separately predict threat and opportunity. Response models are estimated among speeches with at least one clear response and separately predict regulation and restriction, enablement and investment, training, and compensation. Because the coding is multi-label, predicted frame probabilities are not constrained to sum to one.

\paragraph{Government--opposition models.}
The government--opposition comparison uses grouped-binomial GLMs estimated on country--party--year--status cells, with government status and party-family, year, and country fixed effects. Country fixed effects absorb national baselines, while party-family and year fixed effects account for broad ideological and temporal differences.

Government status was coded from manually checked cabinet records. Confidence-and-supply parties, non-party speeches, and ministerial speeches that could not be linked to a party are excluded. We interpret these estimates descriptively: they show whether government and opposition parties differ after accounting for party family, country, and year, but do not identify the causal effect of entering government.

\begin{table}[t]
\centering
\caption{Corpus and coding overview}
\label{tab:overview}
\begin{tabular}{@{}lr@{}}
\toprule
Quantity & Count \\
\midrule
Substantive speeches & 1,514,950 \\
AI candidate speeches & 19,411 \\
AI-work relevant speeches & 5,317 \\
Country cases & 33 national parliaments \\
Country-party units & 936 \\
Observation period & 2023--April 2026 \\
Human validation observations & 728 \\
\bottomrule
\end{tabular}
\end{table}

\section{Findings}

\subsection{AI-work politics is emerging, but unevenly}

\begin{figure}[t]
\centering
\includegraphics[width=\textwidth]{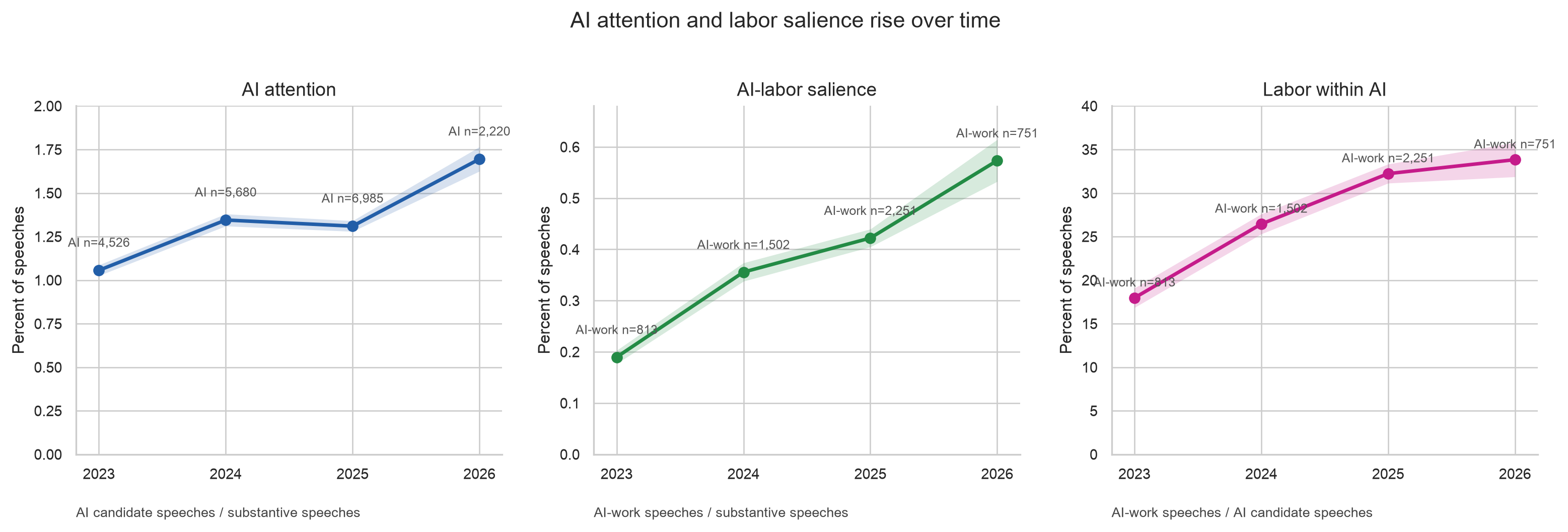}
\caption{AI attention and work salience over time. Panel (a) plots AI candidate speeches as a share of substantive speeches, panel (b) AI-work speeches as a share of substantive speeches, and panel (c) AI-work speeches as a share of AI candidate speeches. The 2026 estimates cover January through April.}
\label{fig:trend}
\end{figure}

AI has entered parliamentary debate, and work is becoming increasingly central to how parliaments make sense of it. AI candidate speeches rise from 1.06 percent of all substantive speech in 2023 to 1.70 percent in 2026, while the share of AI debate connected to work increases from 18.0 to 33.8 percent. Parliaments are therefore both talking more about AI and increasingly linking it to employment, skills, workplace control, productivity, and labor-market change (Figure~\ref{fig:trend}).

The issue nevertheless remains small in absolute terms. Even in 2026, AI-work speech accounts for well below one percent of substantive parliamentary debate. Rapid growth alongside low salience is consistent with an issue still taking shape, as parties begin to define positions before the conflict has become a settled axis of competition.

\begin{figure}[t]
\centering
\begin{subfigure}[t]{0.50\textwidth}
    \centering
    \includegraphics[width=\linewidth]{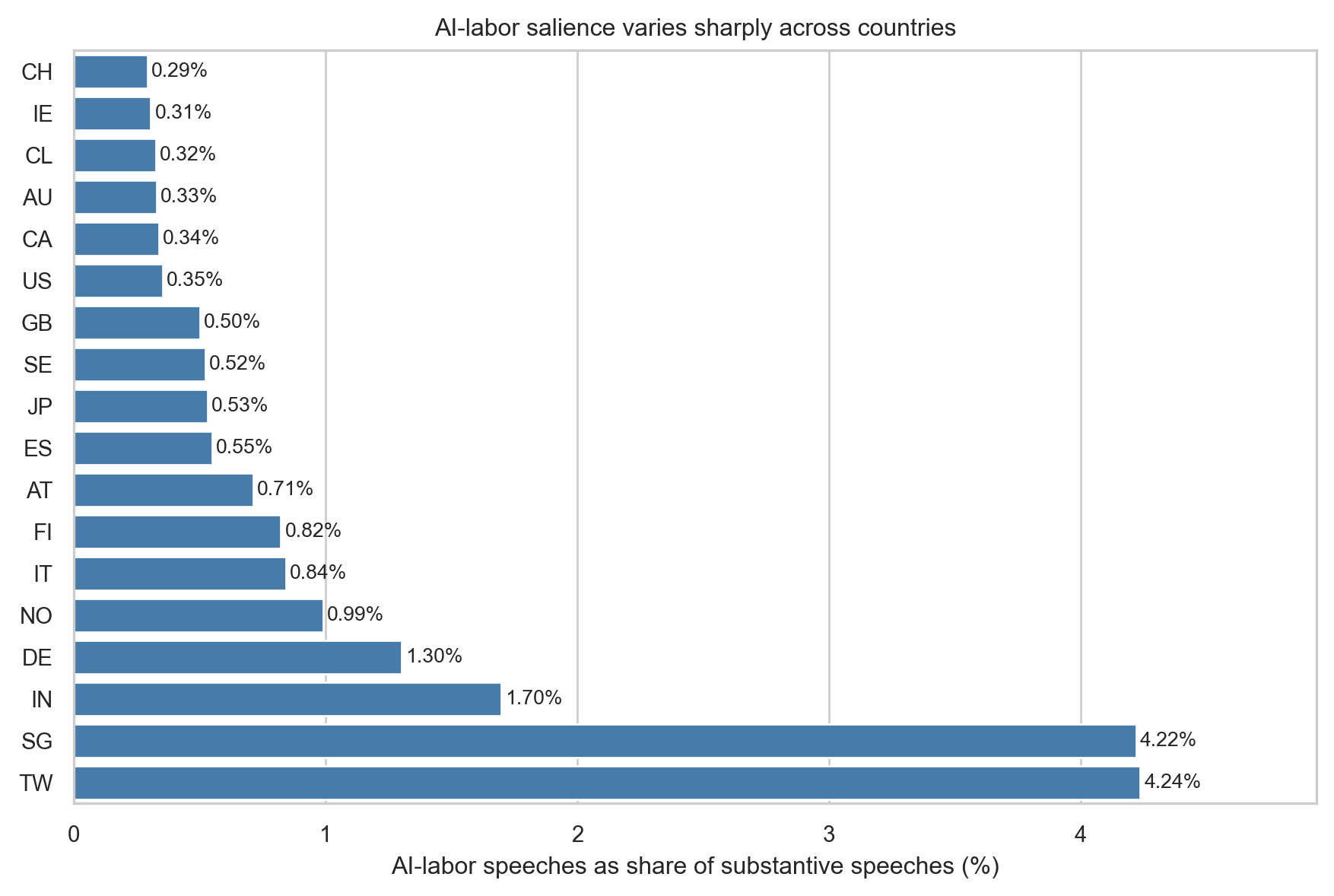}
    \caption{Top country cases}
\end{subfigure}\hfill
\begin{subfigure}[t]{0.47\textwidth}
    \centering
    \includegraphics[width=\linewidth]{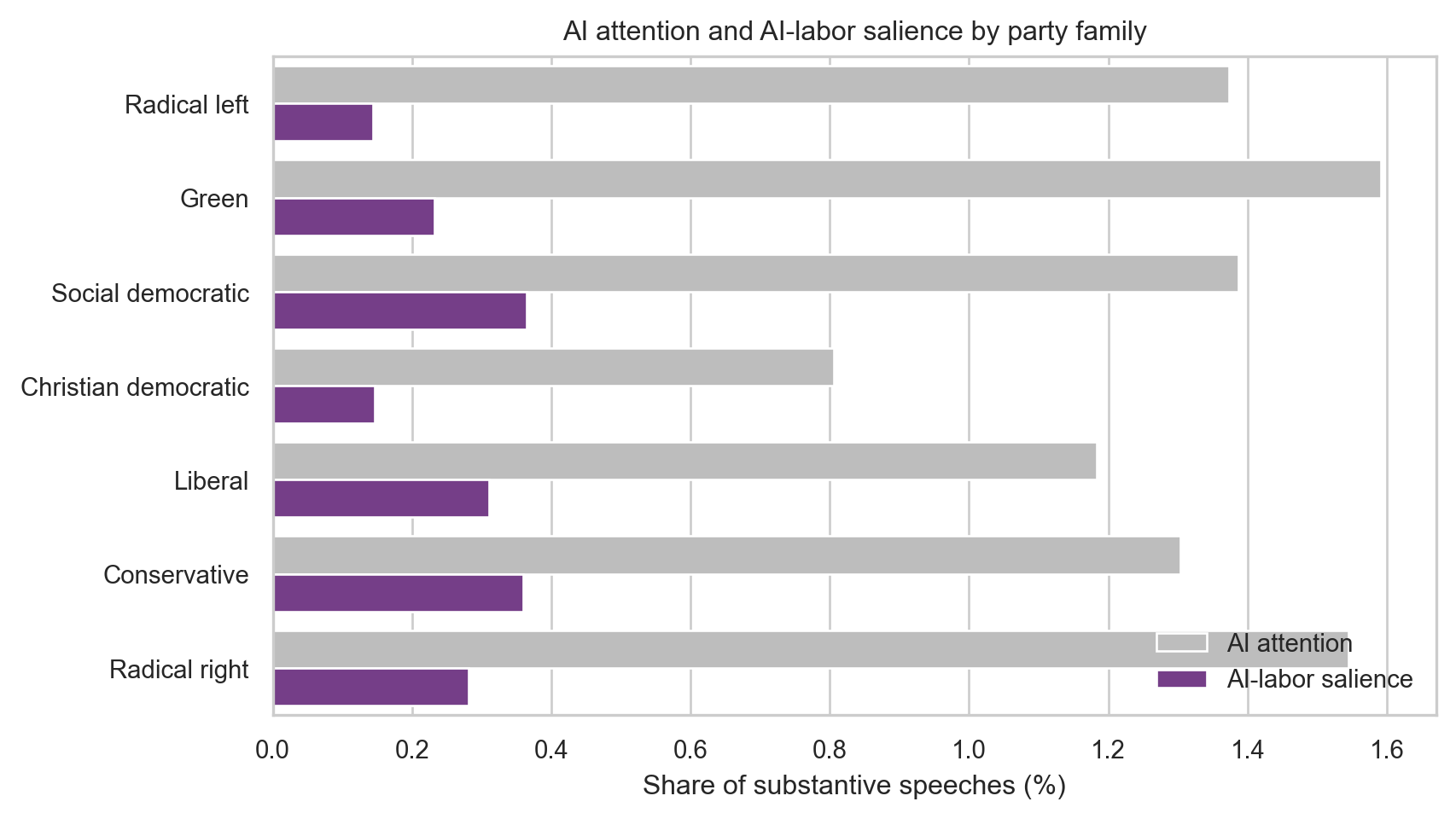}
    \caption{Party families}
\end{subfigure}
\caption{AI-work salience across countries and party families. Panel (a) shows the highest-salience countries among parliaments with at least 1{,}000 substantive speeches; panel (b) compares AI attention and AI-work salience across party families.}
\label{fig:saliencecontext}
\end{figure}

Emergence is also highly uneven across countries. Taiwan and Singapore have the highest AI-work salience, while other relatively high-salience cases are spread across Asia and Europe (Figure~\ref{fig:saliencecontext}). Differences in attention across party families are much smaller, previewing a central pattern in the analysis: parties differ less in whether they discuss AI and work than in the political meaning they give it.

\begin{figure}[t]
\centering
\includegraphics[width=\textwidth]{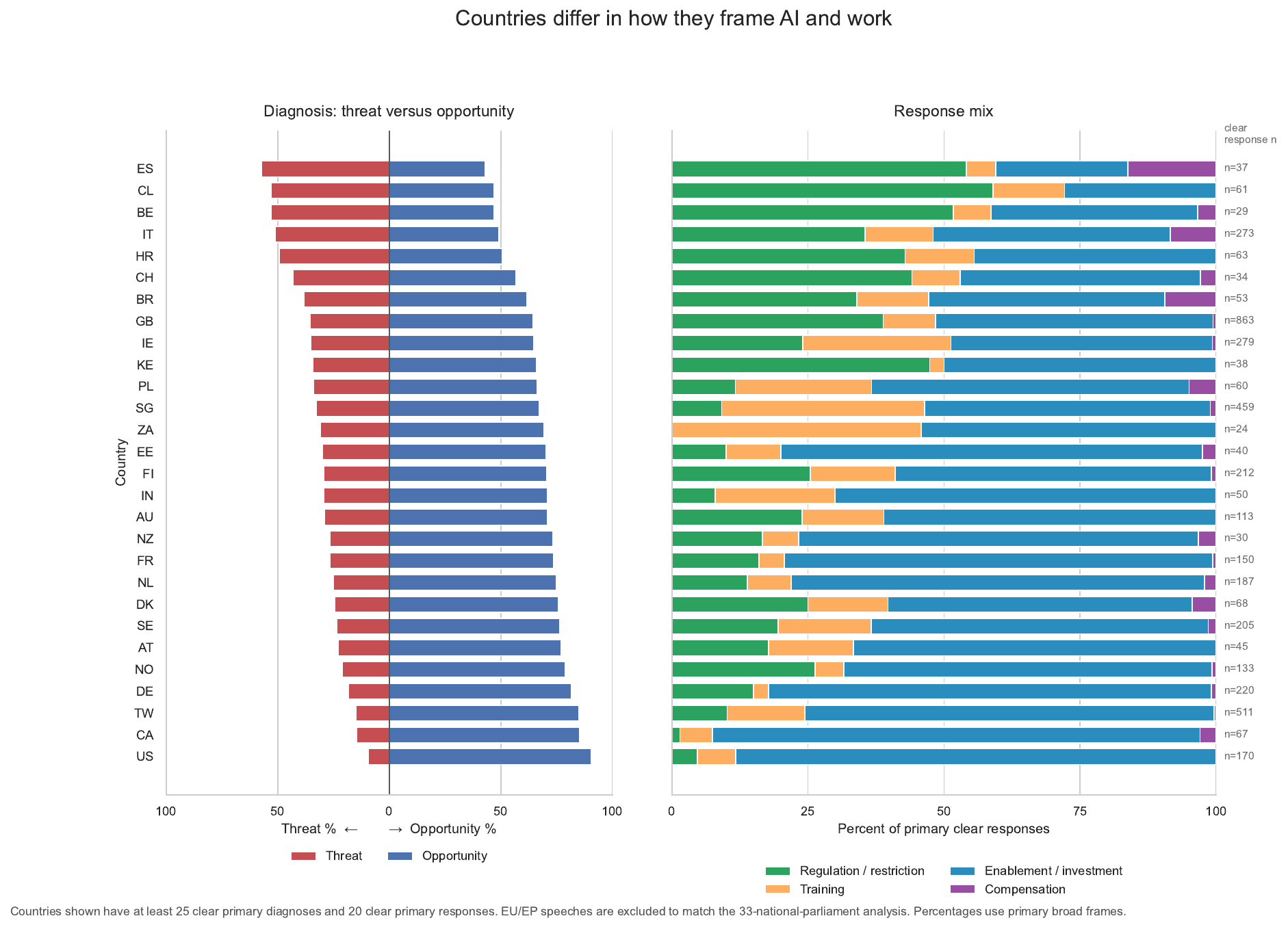}
\caption{Country differences in AI-work diagnostic and response frames. The left panel shows primary diagnostic balance; the right shows the primary response mix. Countries are included with at least 25 clear primary diagnoses and 20 clear primary responses.}
\label{fig:countryframes}
\end{figure}

Country profiles likewise vary in diagnosis and response (Figure~\ref{fig:countryframes}). Spain, Chile, Belgium, and Italy are among the more threat-oriented cases, whereas the United States, Canada, Taiwan, and Germany lean strongly toward opportunity. Responses differ in degree, but around a common asymmetry: enablement and investment is the largest primary response in most countries, while compensation is marginal almost everywhere. We treat these profiles descriptively and turn to party families for the main comparative analysis, accounting for country-specific baselines.

\subsection{Party conflict begins with diagnosis}

\begin{figure}[t]
\centering
\includegraphics[width=0.74\textwidth]{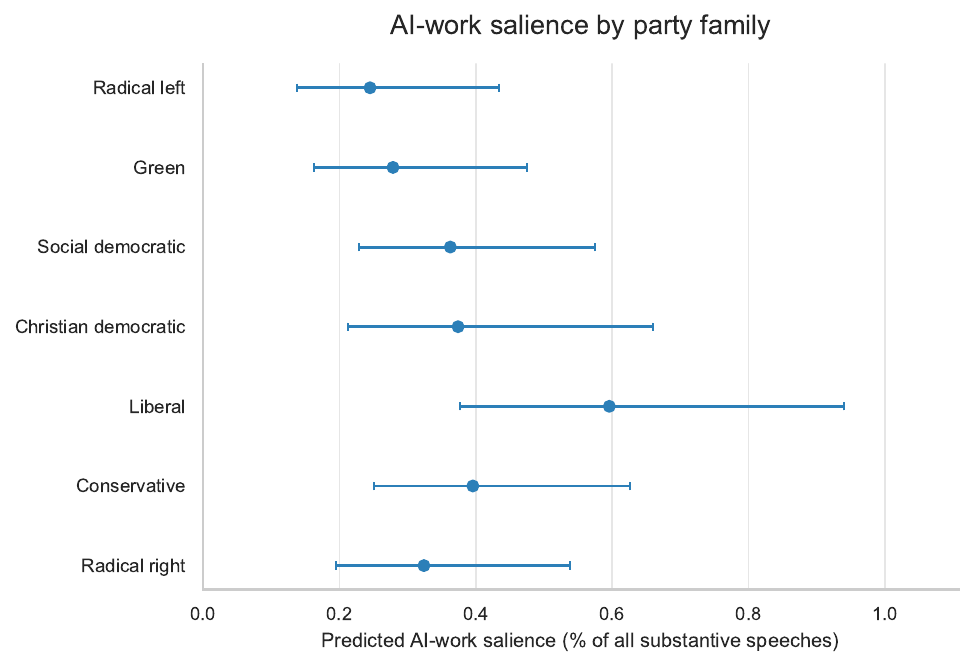}
\caption{AI-work salience by party family. Points report predicted shares from grouped-binomial multilevel models with party-family and year fixed effects and random intercepts for country and country-party. Lines are 95\% intervals.}
\label{fig:partysalience}
\end{figure}

The party-family models confirm that differences in attention are modest across party families. Liberal parties have the highest predicted AI-work salience and green and radical-left parties the lowest, but the uncertainty intervals overlap substantially (Figure~\ref{fig:partysalience}). No family clearly owns the issue in terms of attention; the important differences emerge in how AI is constructed once parties address its consequences for work.

\begin{figure}[t]
\centering
\includegraphics[width=\textwidth]{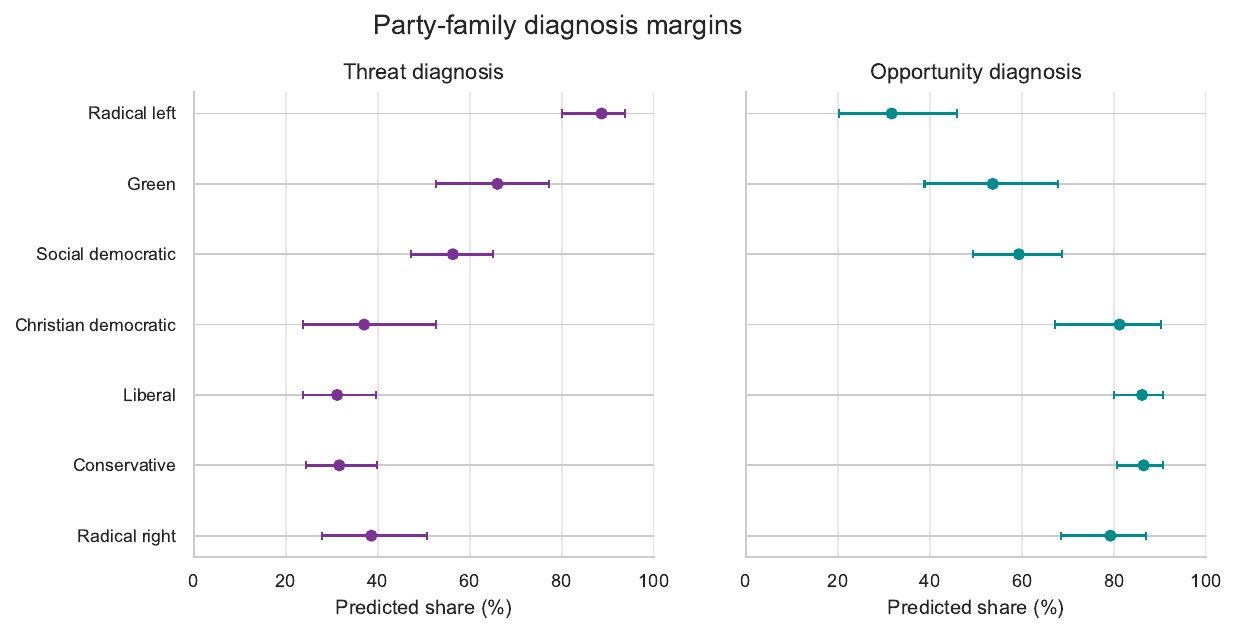}
\caption{Party-family differences in threat and opportunity frames. Points report predicted shares from grouped-binomial multilevel models with party-family and year fixed effects and random intercepts for country and country-party. Lines are 95\% intervals. Because diagnosis is multi-label, shares need not sum to 100\%.}
\label{fig:diagnosis}
\end{figure}

Those differences, however, are pronounced. Roughly nine in ten diagnostically clear radical-left speeches frame AI as a threat to work, while fewer than one in three invoke opportunity (Figure~\ref{fig:diagnosis}). Greens also lean toward threat, though less decisively. Liberal, conservative, Christian-democratic, and radical-right parties cluster at the other end, invoking opportunity in roughly 80--86 percent of diagnostically clear speeches and threat in around one-third. Social democrats occupy the middle, making threat and opportunity visible at similar rates.

 Radical-right parties do not emerge as entrepreneurs of AI-related labor market anxiety. Their opportunity framing closely resembles that of the mainstream right, while threat is only modestly more common. The pattern fits a repertoire logic in which technological vulnerability is filtered through political languages of national competitiveness, technological strength, and the danger of falling behind.

\begin{figure}[t]
\centering
\includegraphics[width=\textwidth]{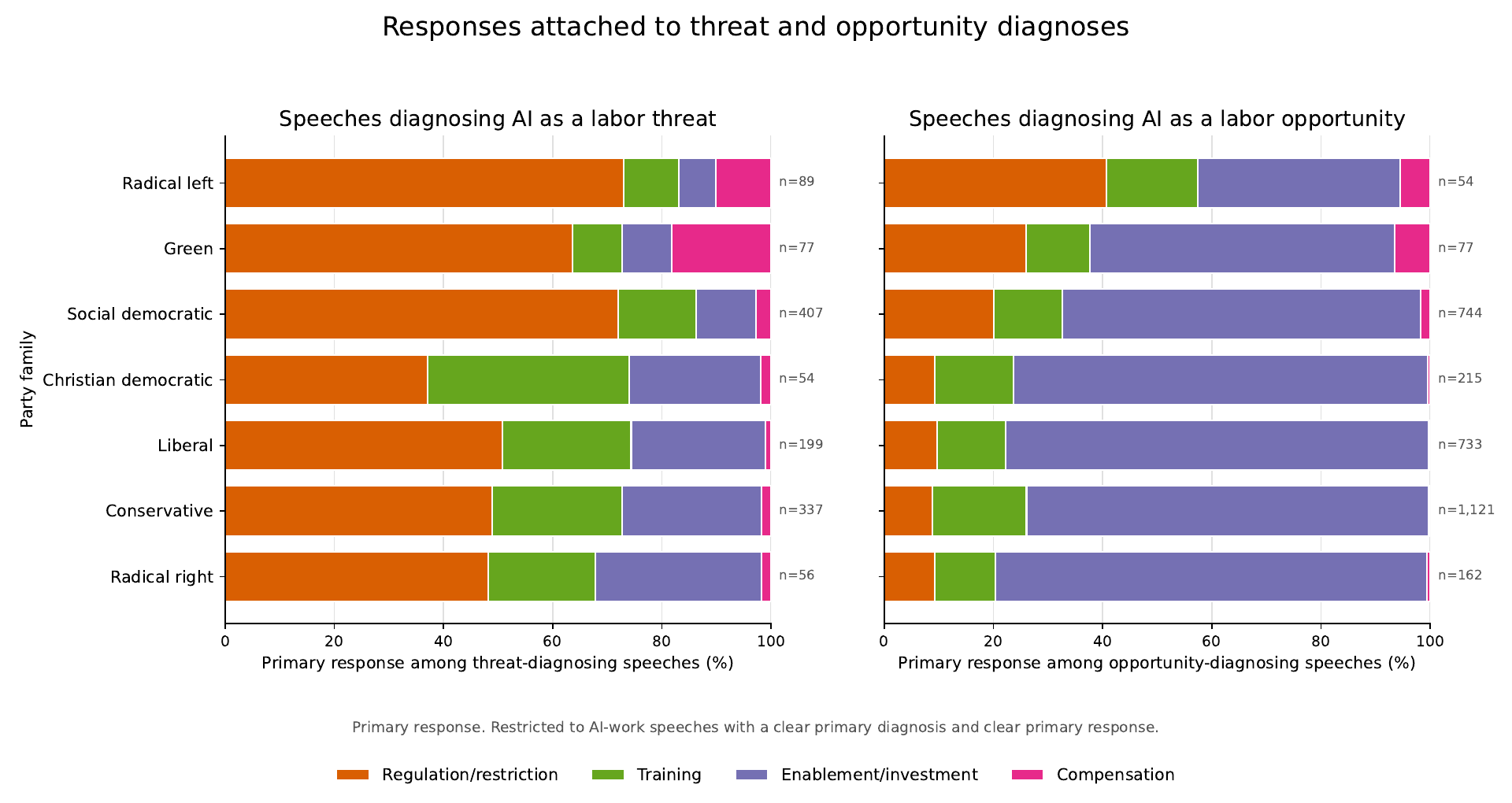}
\caption{Responses attached to threat and opportunity diagnoses. Bars show the primary response among speeches with a clear primary diagnosis and response. The comparison is descriptive and based on mutually exclusive primary frames.}
\label{fig:conditionalresponse}
\end{figure}

Figure~\ref{fig:conditionalresponse} suggests that diagnosis largely structures the response.  Comparing the primary responses attached to threat and opportunity diagnoses across party families, threat-diagnosing speeches lean strongly toward regulation and restriction, while opportunity-diagnosing speeches overwhelmingly favor enablement and investment.

This pattern clarifies the social-democratic position. Threat-diagnosing speeches generally attach regulatory responses, whereas opportunity-diagnosing speeches overwhelmingly favor enablement. The family's intermediate position therefore reflects two substantial streams: one that makes AI's consequences for workers politically problematic and another centered on adoption, investment, and modernization.

More broadly, diagnosis appears as the central site of conflict. The radical left consistently makes AI's labor-market consequences politically problematic, greens do so frequently, and social democrats much of the time; parties of the right overwhelmingly foreground opportunity, productivity, and modernization. Much of the emerging struggle therefore concerns which consequences are made visible and politically actionable in the first place.

\subsection{An asymmetric conflict over technological direction}

\begin{figure}[t]
\centering
\includegraphics[width=0.82\textwidth]{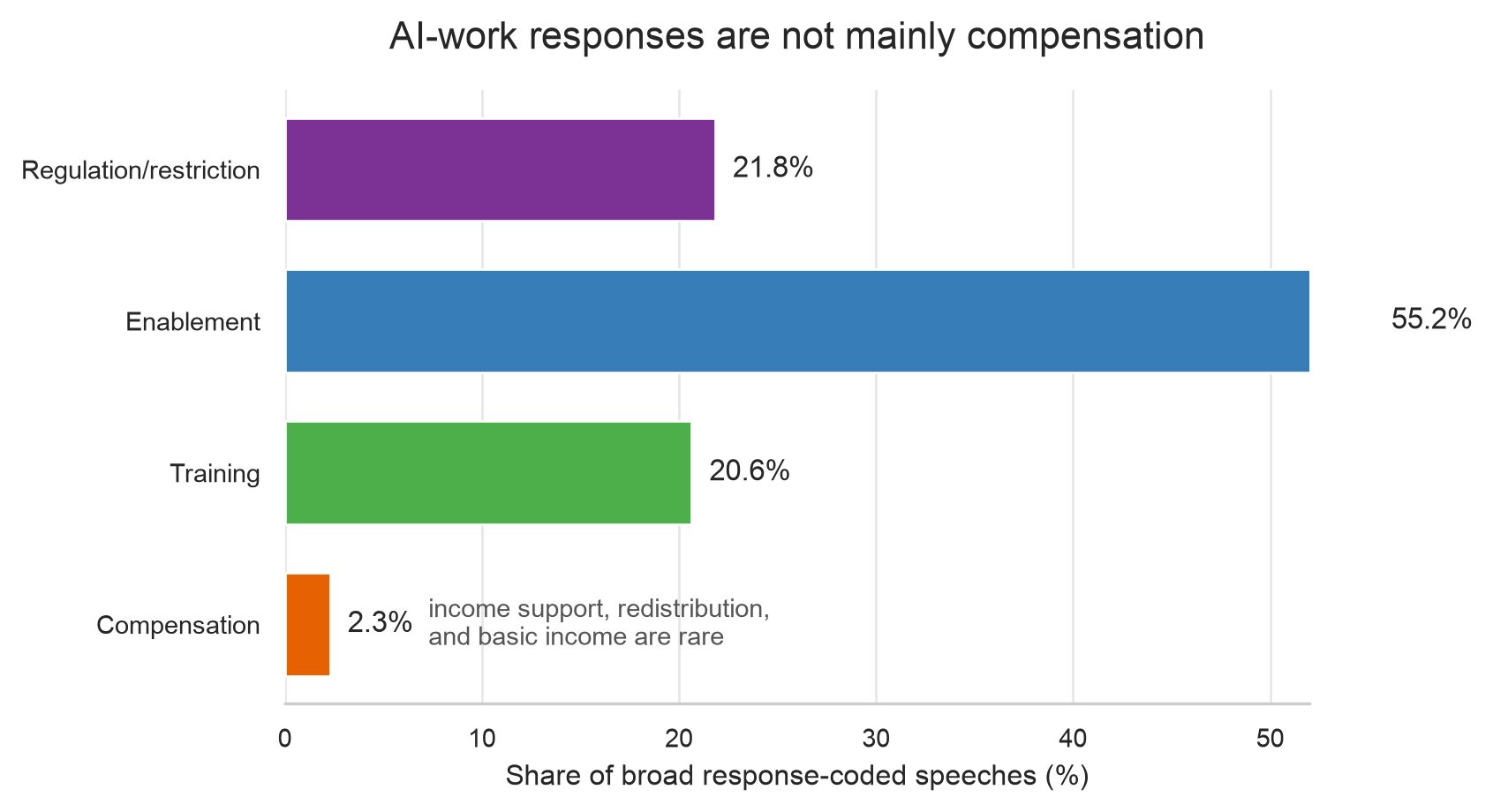}
\caption{Broad response frames in AI-work debate. Bars show the distribution of broad response-frame mentions, counting primary and secondary tags.}
\label{fig:broadresponse}
\end{figure}

Across all broad response-frame mentions, enablement and investment account for 55.2 percent, regulation and restriction for 21.8 percent, training for 20.6 percent, and compensation for only 2.3 percent (Figure~\ref{fig:broadresponse}). Parliamentary debate is thus concentrated on what should happen while technological change unfolds: accelerating adoption, preparing workers and institutions, or governing deployment.

\begin{figure}[t]
\centering
\includegraphics[width=\textwidth]{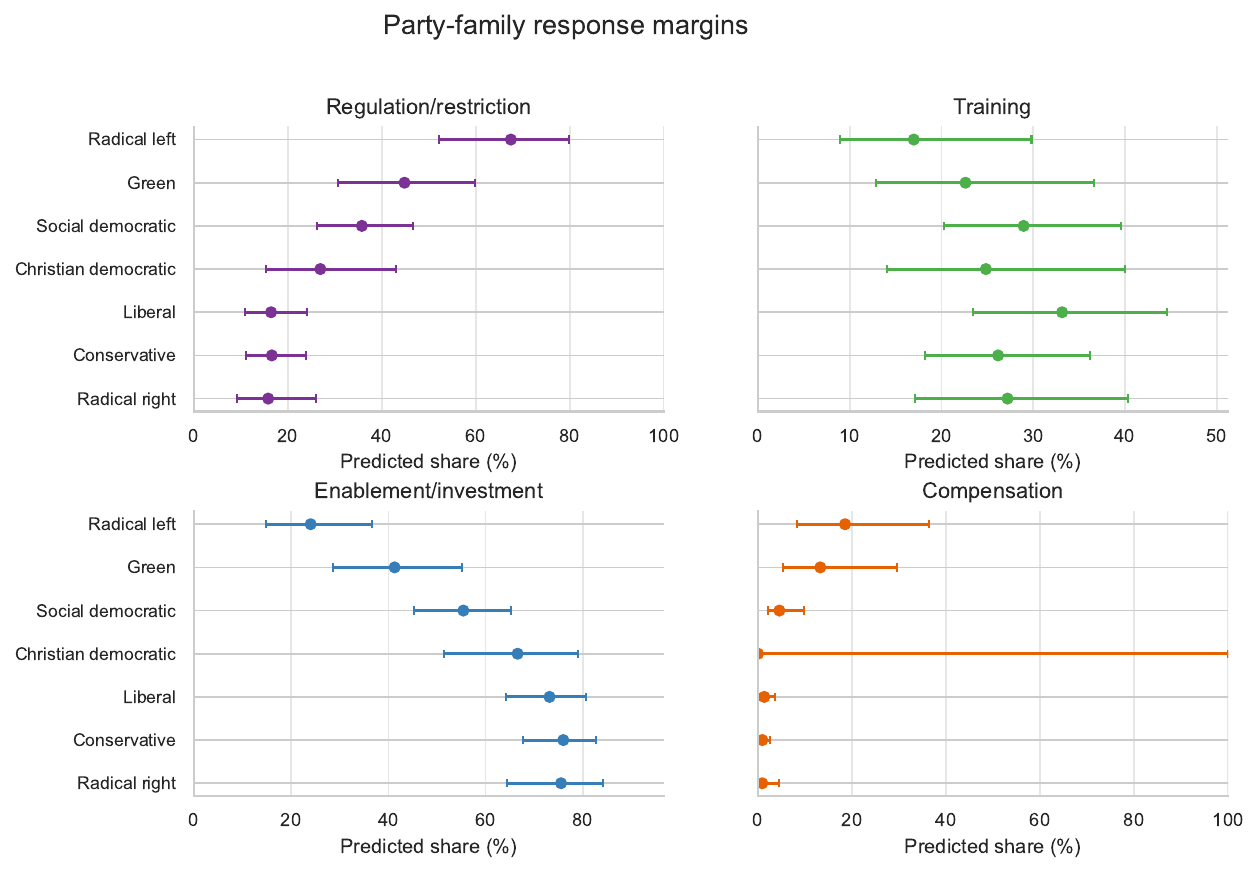}
\caption{Party-family differences in response frames. Points report predicted shares from grouped-binomial multilevel models with party-family and year fixed effects and random intercepts for country and country-party. Lines are 95\% intervals. Because responses are multi-label, shares need not sum to 100\%.}
\label{fig:response}
\end{figure}

The aggregate imbalance conceals a clear partisan structure (Figure~\ref{fig:response}). Enablement and investment dominates among conservative, radical-right, liberal, and Christian-democratic parties and remains the largest response among social democrats. Regulation and restriction follows the opposite gradient, forming the dominant radical-left response and standing roughly level with enablement among greens. The radical left is consequently the only family that consistently combines an overwhelming threat diagnosis with a predominantly governance-oriented response.

\begin{figure}[t]
\centering
\includegraphics[width=0.72\textwidth]{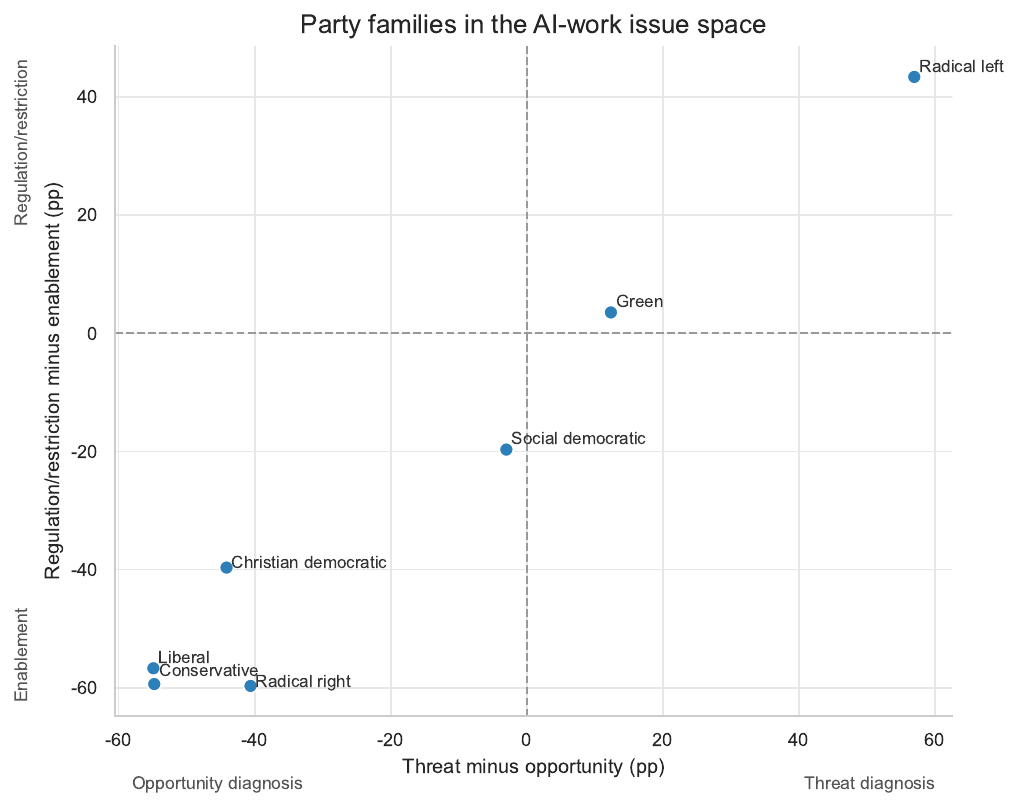}
\caption{Party families in the AI-work issue space. Coordinates are descriptive transformations of the multilevel margins in Figures~\ref{fig:diagnosis} and~\ref{fig:response}: threat minus opportunity on the horizontal axis, and regulation and restriction minus enablement and investment on the vertical axis.}
\label{fig:partyspace}
\end{figure}

Figure~\ref{fig:partyspace} summarizes this structure by combining diagnostic and response margins. Conservatives, liberals, Christian democrats, and the radical right cluster in the adoption-oriented region, joining opportunity diagnoses to enabling responses. Social democrats sit near the diagnostic midpoint but remain on the enablement side in aggregate response. The radical left occupies the opposite end of the space, while greens fall between its contestatory position and the broader adoption-oriented mainstream.

Party families, in other words, line up along a single dimension, from those who treat AI primarily as an opportunity to be enabled to those who treat it primarily as a threat to be regulated. We refer to this as the \emph{enablement–regulation axis}: contestation over whether public authority should accelerate and invest in technological change or condition and constrain it, before its distributive consequences are settled. 


Compensation remains rare throughout the party system despite a broad definition that includes income protection, social protection, redistribution, reduced working time, and basic income. It is somewhat more visible among the radical left, though still clearly secondary to regulation and restriction. Stable constituencies of AI losers may not yet have formed at sufficient scale to organize conflict around post-disruption repair; party competition is already developing further upstream, over how a still-unsettled transformation should proceed.

Training is invoked at broadly similar rates across most party families. This may be explained by its capacity to serve several projects: building the capacity needed for adoption, helping workers navigate disruption, or developing national skills and research capacity. Its political portability allows parties to support it whether they favor accelerating technological change or governing it.

The broad ordering is robust to a narrower definition of AI-work relevance. In direct-only analyses, compensation rises somewhat but remains a small minority response, while enablement and training remain much more prominent. Excluding Singapore and Taiwan, the two highest-salience cases, also leaves the pattern unchanged (Appendix~A).

\subsection{Government parties are more adoption-oriented}

\begin{figure}[t]
\centering
\includegraphics[width=0.9\textwidth]{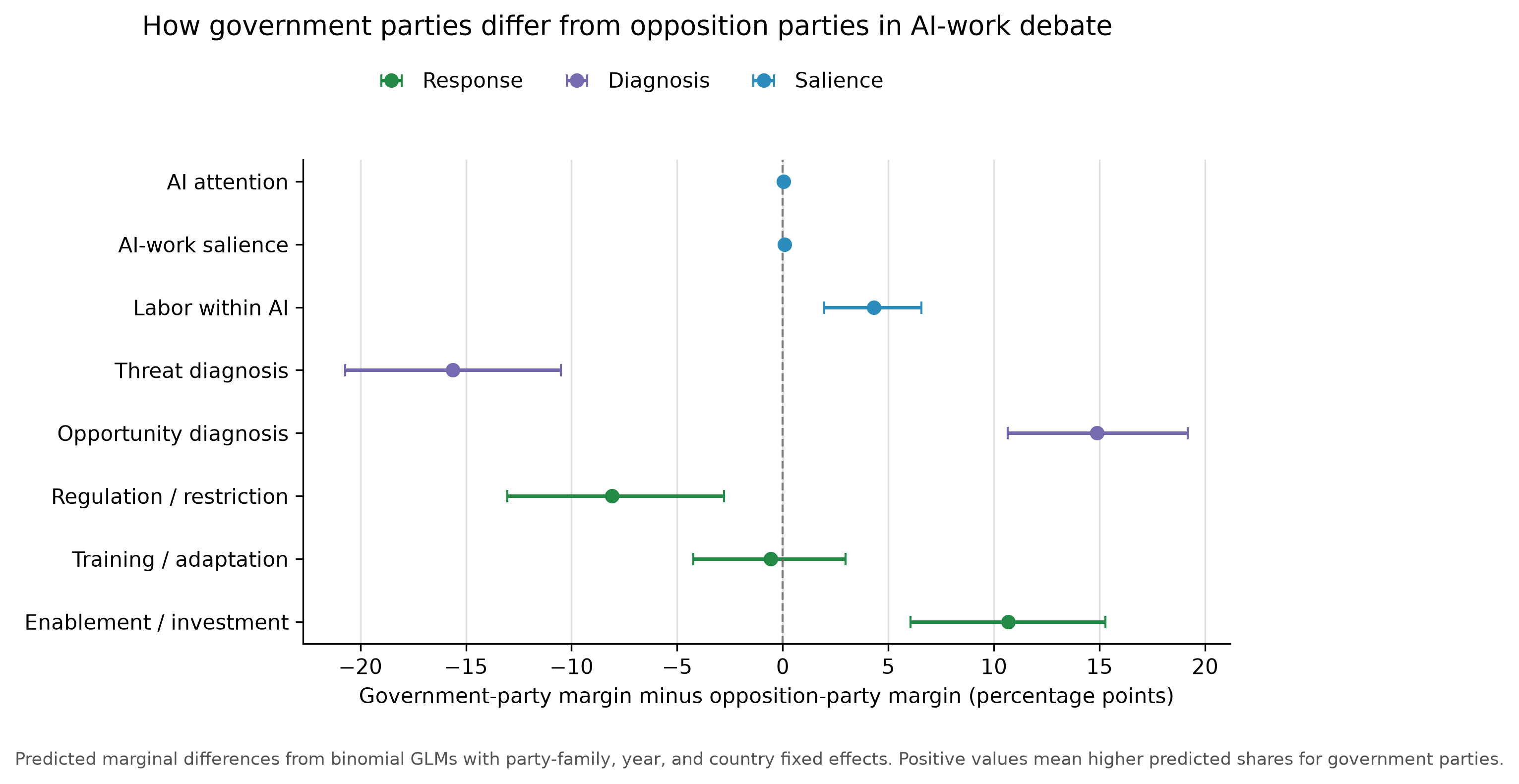}
\caption{Government and opposition differences in AI-work framing. Points report predicted shares from grouped-binomial GLMs with government status, party-family, year, and country fixed effects. Lines are 95\% confidence intervals.}
\label{fig:role}
\end{figure}

Government status adds a secondary gradient: net of party family, country, and year, government-party speech is somewhat more likely to make AI and work salient, frame AI as an opportunity, and emphasize enablement and investment; opposition parties are more threat-oriented and more likely to favor regulation or restriction (Figure~\ref{fig:role}). Training differs little by institutional position.

While the comparison is associational and does not identify the effect of entering office, it is nevertheless consistent with the pressures of governing responsibility, which make economic performance, procurement, public-sector modernization, and national strategy more salient. These pressures reinforce the adoption-oriented center of gravity without accounting for the larger ideological structure separating the contestatory left from the rest of the party system.

\subsection{Exploratory analysis}
\begin{figure}[t]
\centering
\includegraphics[width=\textwidth]{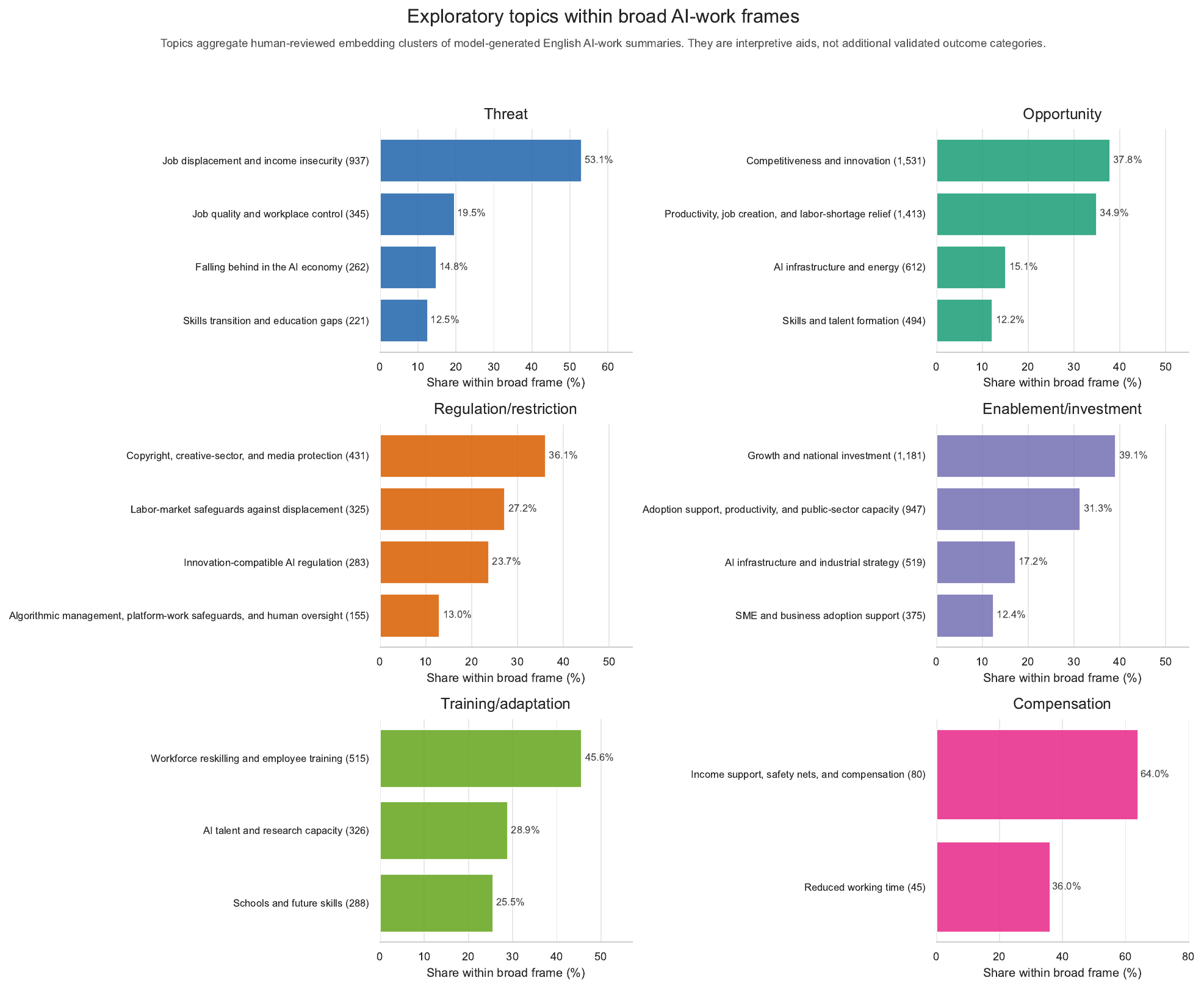}
\caption{Exploratory topics within broad AI-work frames. Bars show topic shares within each validated broad frame. Topics aggregate human-reviewed embedding clusters and are used as interpretive aids rather than validated outcome categories.}
\label{fig:contentclusters}
\end{figure}
Figure~\ref{fig:contentclusters} opens up the broad frames validated above, showing what parliamentarians are actually talking about within each. The picture that emerges is one of concentration within diagnoses and fragmentation within responses.

On the diagnostic side, threat discourse has a dominant center of gravity: job displacement and income insecurity alone account for 53.1 percent of threat content, more than the next three topics combined. Workplace control (19.5 percent), falling behind in the AI economy (14.8 percent), and skills gaps (12.5 percent) form a secondary tier. Opportunity discourse is more evenly split between two leading topics, competitiveness and innovation (37.8 percent) and productivity, job creation, and labor-shortage relief (34.9 percent), with infrastructure and skills formation trailing well behind. Threat, in other words, is largely one story about disruption to income and livelihoods; opportunity is two stories, one about national competitiveness and one about firm-level productivity, told in roughly equal measure.

The response frames are structured quite differently. Regulation and restriction is led not by algorithmic management, the topic most associated with AI-specific workplace governance, but by copyright, creative-sector, and media protection (36.1 percent) --- a legacy concern about intellectual property and creative labor that generative AI has revived rather than invented. Labor-market safeguards against displacement (27.2 percent) and innovation-compatible regulation (23.7 percent) follow, while algorithmic management, platform-work safeguards, and human oversight --- the topic closest to the surveillance and workplace-control concerns raised on the threat side --- is the smallest category, at just 13.0 percent. 

Enablement is dominated by macro-level investment: growth and national investment (39.1 percent) and adoption support, productivity, and public-sector capacity (31.3 percent) together account for seven in ten enablement mentions, with AI infrastructure and industrial strategy (17.2 percent) and SME adoption support (12.4 percent) trailing. Enablement, in short, is directed at least as much toward building state and public-sector capacity as toward  toward supporting private firms --- public authority is mobilized to expand the state's own technological reach, not only to clear the way for business adoption. Training divides across employee reskilling (45.6 percent), AI talent and research capacity (28.9 percent), and schools and future skills (25.5 percent), consistent with its earlier-noted ability to serve adoption, adaptation, and long-run human-capital investment simultaneously.

Compensation, the rarest broad frame, is also the narrowest in content: nearly all of it (64.0 percent) concerns income support and safety nets, with reduced working time (36.0 percent) as the only substantial alternative. With just 125 mentions across the two topics combined, compensation politics is not merely underused relative to the other response frames; the little that exists is also thin substantively, largely restating conventional welfare-state instruments rather than developing AI-specific remedies such as robot taxes or productivity-sharing schemes.

Table~\ref{tab:examples} illustrates these patterns with representative claims from the corpus. Regulation examples span the range documented above, from workplace-oriented safeguards to the copyright and creative-sector protections that now form its largest single component; enablement, training, and compensation examples show the same response frame doing different work depending on which party and context invokes it.

\begin{table}[t]
\centering
\caption{Illustrative AI-labor claims in parliamentary debate}
\label{tab:examples}
\scriptsize
\begin{tabularx}{\textwidth}{@{}p{0.18\textwidth}p{0.18\textwidth}p{0.17\textwidth}X@{}}
\toprule
Claim type & Party / country & Coded response & Paraphrased content \\
\midrule
Govern work: regulation & Die Linke (DE, 2025) & regulation restriction & Frames AI as a risk to workers and argues for public rules or constraints on deployment. \\
Govern work: restrictions & Australian Labor Party (AU, 2026) & regulation restriction & Links AI to risks in employment or creative work and calls for tighter limits or safeguards. \\
Adoption and productivity & Liberal (CA, 2025) & enablement investment & Connects public AI investment to innovation capacity and the creation of future high-value jobs. \\
Training and adaptation & Labour (GB, 2025) & training adaptation & Acknowledges AI displacement while emphasizing skills, transition support, and free AI training for workers. \\
Compensation & NDP (CA, 2025) & compensation & Argues that AI can replace workers and links the response to employment insurance and income supports. \\
\bottomrule
\end{tabularx}
\end{table}

Together, these patterns sharpen the frame-level findings reported above. The broad categories validated through human coding conceal real heterogeneity beneath them: threat discourse is concentrated on displacement, opportunity discourse is split between competitiveness and productivity, regulation is currently more about copyright than about algorithmic control at work, and enablement leans as heavily on public-sector capacity as on firm support. The distinction that structures the party conflict documented above --- enabling adoption versus regulating it --- is therefore not a single substantive fight, but an umbrella covering several more specific ones, each still in the process of being politically defined.

This heterogeneity also helps explain a puzzle in the party-family results reported above. Enablement is not a single position: growth, national investment, and industrial strategy account for more of it than firm-level adoption support, giving enablement a public-investment register that is ideologically available to the left as much as the right. This may be why social democrats and greens, despite their threat-oriented diagnoses, still register substantial enablement in their response mix.

\section{Discussion and conclusion}
What kind of politics is emerging around AI's labor-market impact? Across 33 national parliaments and more than 1.5 million substantive speeches, we find that it is not, so far, a politics of compensation: compensation accounts for just 2.3 percent of response-frame mentions. Instead, the real conflict runs along what we call an \emph{enablement--regulation axis}: contestation over whether public authority should accelerate and invest in AI's adoption or condition and constrain it. 

Most party families---the mainstream right, the radical right, and, on balance, social democrats---sit on the enabling side of this axis, treating AI as an opportunity and a transformation for which workers, firms, and states should prepare. The radical left stands apart, overwhelmingly constructing AI's consequences for work as threats and favoring regulation and restriction over enablement. Greens lean the same way on diagnosis, albeit less decisively, and split their response evenly between regulation and enablement. The resulting conflict is highly asymmetric: most of the party system seeks to enable technological change, while a smaller current, anchored by the radical left, seeks to instead govern its direction.

This asymmetry is not only about which parties want more or less state action; it is about which consequences get made visible in the first place. Speeches diagnosing threat tend toward regulation and restriction, while those diagnosing opportunity overwhelmingly favor enablement and investment. Framing AI as a threat or an opportunity is therefore not a preliminary to the political conflict---it is a substantial part of it. The radical left goes furthest in this respect, more than any other family pairing a threat diagnosis with a regulation-and-restriction response rather than settling for compensation alone. Greens combine a similar diagnosis with greater openness to enablement, splitting their response evenly rather than favoring restriction outright, which places them between radical-left contestation and the adoption-oriented mainstream. Social democracy's position instead reflects a familiar approach to structural economic change: the broad transformation is accepted, politics manages the transition, and regulation is reserved for its most damaging consequences \citep{GingrichHausermann2015, AbouChadiHausermannMittereggerMosimannWagner2025}. On AI, this leaves the strongest parliamentary challenge to the radical left rather than to the historically dominant party family of organized labor.

Although automation exposure has repeatedly been linked to radical-right support, this party family is not politicizing AI as a threat to workers, but sits alongside the mainstream right in both diagnosis and response, emphasizing opportunity, adoption, competitiveness, and national capacity. This fits a repertoire logic: national competition, sovereignty, and the danger of falling behind make technological adoption a comfortable frame for the radical right, while workplace regulation and institutionalized labor protection fit less easily with its established repertoire \citep{BorweinBonikowskiLoewen2024, enggist2022radical}. More broadly, this speaks to theories of new-issue formation: because AI's material consequences have not stabilized, parties are free to connect it to existing conflicts and reputational resources---national modernization for the mainstream right, worker vulnerability for the radical left, managed transition for social democrats. The same technology is becoming several different political objects at once, depending on who is describing it.

The current structure of parliamentary supply may also leave some forms of technological insecurity weakly represented. Demand-side research finds support for slowing technological change, regulating its effects, and adopting work-preserving interventions \citep{GallegoKuoManzanoFernandezAlbertos2022, heinrich2025self, Chueri2026}, whereas compensation and decommodifying responses remain almost absent from the parliamentary record, and sustained contestation of technological direction is concentrated in a relatively small part of the party system. While our evidence cannot establish a direct representation gap, since demand- and supply-side studies observe different populations, countries, and periods, it nonetheless leaves open the possibility that public unease about AI is outpacing the positions parties have so far built around it.

Institutional position adds a secondary gradient. Government parties are more likely than opposition parties to frame AI as an opportunity and emphasize enablement and investment, net of party family, country, and year. The analysis does not identify the causal effect of entering office, but the pattern is consistent with the pressures of governing responsibility: national strategy, economic performance, procurement, and public-sector modernization all encourage parties to present AI as a governable opportunity. Government status thus reinforces the adoption-oriented center of gravity without accounting for the broader ideological structure.

Several limitations qualify the argument. Parliamentary speech does not directly measure enacted policy, whose formation also depends on coalition bargaining, veto institutions, bureaucracies, courts, unions, employers, and organized interests. The study captures an unusually early period, which makes it possible to observe conflict before AI's labor-market effects have stabilized but also means that the emerging structure may change rapidly: visible displacement could strengthen compensation politics, deepen contestation on the left, or create new opportunities for radical-right mobilization. The country sample is broad but determined by the availability of official, recoverable speech-level records, and differences in parliamentary institutions and transcript practices warrant caution in interpreting national rankings. Finally, the models map associations rather than causal effects of ideology or government participation.

Future research should connect parliamentary framing to enacted policy, examine how unions, employers, technology firms, and social movements shape party positions, and explain cross-national variation in the balance between adoption and contestation. The most important questions are temporal: will visible displacement produce the familiar politics of compensation, will radical-left and green parties consolidate a broader politics of technological direction, will social democracy remain committed to managed adaptation, and will the radical right continue to leave technological protection to the left or translate AI-related insecurity into a nationalist and populist politics of its own?

For now, most parties approach AI as a transformation to be enabled and accommodated, while a smaller current centered on the radical left insists that adoption should be governed and tempered. The pace of adoption, authority at work, protections surrounding deployment, and the distribution of technological gains are being shaped while the technology is still developing---the winners and losers of technological change emerge only later, but the politics that helps produce them has already begun.

\section * {Data Availability Statement}
The data and code necessary to reproduce the analyses reported in this article will be made publicly available in an open repository upon publication.

\appendix

\printbibliography

\end{document}